\documentclass{ws-procs975x65}

\begin{document}

\title{PARAMETERIZING THE DECELERATION PARAMETER}
\author{D. PAV\'{O}N$^*$ and I. DURAN}
\address{Departamento de F\'{\i}sica, Universidad Aut\'{o}noma de Barcelona,\\
Bellaterra, 08193, Spain\\
$^*$E-mail: diego.pavon@uab.es}
\author{S. DEL CAMPO and R. HERRERA}
\address{Instituto de F\'{\i}sica, Pontificia Universidad
Cat\'{o}lica de Valpara\'{\i}so, Chile}
\begin{abstract}
We propose and constrain with the latest observational data three
parameterizations of the deceleration parameter, valid from the
matter era to the far future. They are well behaved and do not
diverge at any redshift. On the other hand, they are model
independent in the sense that in constructing them the only
assumption made was that the Universe is homogeneous and isotropic
at large scales.
\end{abstract}


\bodymatter

\section{Introduction}
\label{sec:intro}
The deceleration parameter, $q = -1 -(\dot{H}/H^{2})$, is poorly
known at present. This is why many parameterizations of this key
quantity, such as $q = q_{0} \, + \, q_{1} z$, $q = q_{0} \, + \,
q_{1} z (1 \, + \, z)^{-1}$, $q = q_{1} \, +\, q_{2}z(1+z)^{-2}$,
$q = 1/2 \, + \, q_{1}(1+z)^{-2}$, $q = 1/2 \, + \, (q_{1}z \, +
\, q_{2})(1+z)^{-2}$, and more complex than these, has been
proposed to reconstruct $q(z)$ from observational data (see e.g.
Refs. \refcite{elgaroy}-\refcite{nair}). However, the first
parameterization is appropriate for $\mid z \mid \ll 1$ only, and
the others diverge in the far future (as $z \rightarrow -1$).

Here we propose three model independent parameterizations of
$q(z)$ with two free parameters only, valid from matter domination
($z \gg 1$) onwards (i.e., up to $z = -1$), based on practical and
theoretical reasons and independent of any cosmological model.
They obey by construction the asymptotic conditions, $q(z \gg 1) =
1/2$, $q(z = -1) = -1$, and a further condition, $dq/dz > 0$,
which is valid at least when $q \rightarrow -1$. The first one
arises because at sufficiently high redshift the Universe was
matter dominated. The other conditions are based on the second law
of thermodynamics when account is made of the entropy of the
causal horizon. The latter dominates over all other entropy
sources\cite{egan} and is proportional to the horizon area, ${\cal
A}$. Then, the second law of thermodynamics\cite{callen} imposes
${\cal A}' \geq 0$ at all times, and ${\cal A}'' \leq 0$ at least
at late times (derivatives are taken with respect to the scale
factor). This translates into $q(z) \geq -1$ (at any redshift),
and that $q \rightarrow -1$ and $dq(z)/dz >0$ as $z \rightarrow
-1$ (see Ref.~\refcite{sird} for details).

\section{Parameterizations}
\label{sec:parm}
Usually one parameterizes a function in any specific interval by
interpolating it between the two end points  of the interval
(modulo one first knows the value taken by the function at these
two points). In actual fact, the parameterizations of $q(z)$
proposed so far have just one fixed point: the asymptotic value at
high redshift ($q$ must converge to $1/2$ when $z \gg 1$). The
other, $q_{0}$, is not in reality a fixed point because the value
of the deceleration parameter at $z = 0$ is not very well known
and it is therefore left free. The parameterizations proposed here
have two fixed points, one at the far past ($z \gg1$), and other
at the far future ($z=-1$). The second fixed point conforms to the
thermodynamical constraints imposed by the second law. We believe
this means a clear advantage over previous parameterizations of
$q(z)$, with just one fixed point. While in the literature it can
be found parameterizations that also fix $q$ at $z = -1$ they do
so arbitrarily, i.e., not grounded on sound physics.

\begin{figure}[t]
\begin{center}
\psfig{file=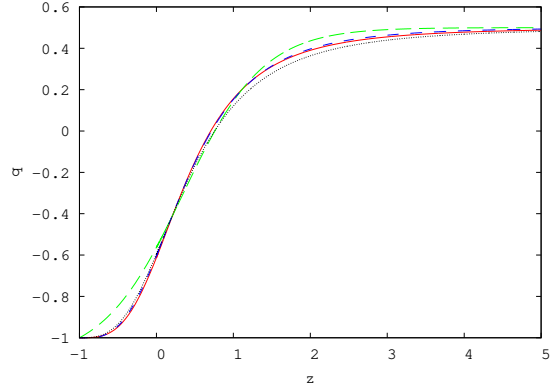,width=3in}
\end{center}
\caption{Deceleration parameters vs redshift. Solid (red), long
dashed (blue) and short dashed (green) lines are for
parameterizations I, II and III respectively. The dotted (black
line) corresponds to the $\Lambda$CDM model which is included for
the sake of comparison. The graphs of parameterizations I and III
practically overlap each other.} \label{aba:fig1}
\end{figure}

We propose three parameterizations of $q(z)$, namely:
\begin{equation}
q(z)
=-1+\frac{3}{2}\left(\frac{(1+z)^{q_{2}}}{q_{1}+(1+z)^{q_{2}}}\right)\,
, \label{eq:parm1}
\end{equation}
\begin{equation}
q(z)=
-\frac{1}{4}\left(3q_{1}+1-3(q_{1}+1)\frac{q_{1}e^{q_{2}(1+z)}-e^{-q_{2}(1+z)}}
{q_{1}e^{q_{2}(1+z)}+e^{-q_{2}(1+z)}}\right) \, , \label{eq:parm2}
\end{equation}
and
\begin{equation}
q(z) = -\frac{1}{4} \, + \, \frac{3}{4} \; \frac{q_{1}
e^{q_{2}\frac{z}{\sqrt{1+z}}}-e^{-q_{2}
\frac{z}{\sqrt{1+z}}}}{q_{1}
e^{q_{2}\frac{z}{\sqrt{1+z}}}+e^{-q_{2}\frac{z}{\sqrt{1+z}}}} \, .
\label{eq:parm3}
\end{equation}
All of them satisfy the conditions stated above. Their two free
parameters, $q_{1}$ and $q_{2}$, were constrained using data from
SN Ia (557 data points), BAO combined with CMB (7 data points) and
the history of the Hubble factor (24 data points). Table
\ref{aba:tbl1} shows their best fit values and their 1$\sigma$
confidence levels. Likewise, figure \ref{aba:fig1} depicts the
corresponding $q(z)$ graphs and that of the $\Lambda$CDM model
fitted to the same sets of data. Details can be found in Ref.
\refcite{sird}.

\begin{table}
\tbl{Best fit values of the free parameters of the three
parameterizations.} {\begin{tabular}{p{1.2 cm} p{2.6 cm} p{2.6 cm}
p{2.4 cm}} \toprule
 & Param. I & Param. II & Param. III  \\
 $\; \; q_{1} $ &$\; \, 2.87^{+0.70}_{-0.53} $ & $\;
\, 0.078^{+0.086}_{-0.043}$ & $\; \,0.36^{+0.07}_{-0.08}$  \\
$\; \; q_{2} $ &$\; \, 3.27\pm 0.55 $ & $\; \,
0.95^{+0.23}_{-0.20}$ & $\; \, \, 1.57^{+0.27}_{-0.33}$ \\
 \botrule
\end{tabular}}
\label{aba:tbl1}
\end{table}

Table \ref{aba:tbl2} gives Hubble's constant, $H_{0}$ (in
km/s/Mpc), the age of the Universe, $t_{0}$  (in Gyr), the
deceleration parameter, $q_{0}$, and the redshift, $z_{t}$, of the
transition deceleration-acceleration predicted for the  three
parameterizations, and the flat $\Lambda$CDM model. The latter is
included for comparison.
\begin{table}
\tbl{Predicted value of some key cosmological parameters.}
     {\begin{tabular}{p{1.2 cm} p{2.6 cm} p{2.6 cm} p{2.4 cm} p{2.7 cm}}
\toprule
 & Param. I & Param. II & Param. III &$\; \;\;\;\;  \Lambda$CDM \\
 $\; \; H_{0} $ &$\; \, 70.5^{+1.5}_{-1.6} $ & $\; \,
70.4 \pm 1.6$ & $\; \, 70.5^{+1.4}_{-1.6}$& $\; \; \; \; 70.2 \pm
1.4 $ \\
$\; \; t_{0} $ &$\; \, 13.6 \pm 0.5$ & $\; \, 13.7\pm 0.4$
& $\; \, \, 13.6\pm 0.2$& $\; \; \; \; 13.4\pm 0.1$ \\
 $\; \; q_{0} $ &$\; -0.61^{+0.06}_{-0.07}$ & $\; -0.56^{+0.35}_{-0.22}$& $\;-0.60\pm 0.06$ &
 $\; \, -0.60 \pm0.03$ \\
$\; \; z_{t}$ & $\; \; \; 0.71^{+0.14}_{-0.17}$ &$\;  \; \;\,
0.77^{+0.52}_{-0.57} $ & $\;  \; \;\, 0.72^{+0.27}_{-0.21}$ &
$\; \; \; \; \; 0.76 \pm 0.05$ \\
\botrule
\end{tabular}}
\label{aba:tbl2}
\end{table}
\section{Conclusions}
The three parameterizations proposed here rest solely on the
assumptions that the Universe is homogeneous and isotropic at
large scales and on the second law of thermodynamics. They agree
very well with each other (especially the first and third) and
with the $\Lambda$CDM model. Likewise, they also concord with the
measurements reported by Daly {\em et al.}, Ref.~\refcite{daly},
in the redshift interval $0 <z < 1$. It is worthy of mention that,
as argued in Ref.~\refcite{sird}, our restriction to spatially
flat models ($k = 0$) is, in reality, very light and well
justified.

\section*{Acknowledgments}
This work was partially supported by the Chilean grant FONDECYT
N$_{0}$ 1110230.


\end{document}